\documentclass[prb,superscriptaddress,twocolumn,showpacs,preprintnumbers,
amsmath,amssymb]{revtex4}

\usepackage{graphicx} % Include figure files
\usepackage{dcolumn} % Align table columns on decimal point
\usepackage{bm}
\usepackage{txfonts}
\usepackage{mathrsfs}
\usepackage{feynmf}
\usepackage{comment}

\usepackage[]{psfrag}

\psfrag{tyldu}{$\tilde{u}$}
\psfrag{deltal}{$\delta$}
\psfrag{logdelt}{$\log (\delta-\delta_c)$}
\psfrag{logphi}{$\log (\phi_0)$}
\psfrag{dgdc}{$\frac{\delta_G-\delta_c}{\delta_c}$}
\psfrag{Temp}{$T$}
\psfrag{etaom}{$\eta_\omega$}

\begin{document}

%\preprint{ }

\title{Renormalization group for phases with broken discrete symmetry near 
 quantum critical points}

\author{P. Jakubczyk}
\email{p.jakubczyk@fkf.mpg.de}
\affiliation{Max-Planck-Institute for Solid State Research,
Heisenbergstr.~1, D-70569 Stuttgart, Germany}
\affiliation{Institute for Theoretical Physics, Warsaw University, 
 Ho\.za 69, 00-681 Warsaw, Poland}
\author{P. Strack}
%\email{p.strack@fkf.mpg.de}
\affiliation{Max-Planck-Institute for Solid State Research,
Heisenbergstr.~1, D-70569 Stuttgart, Germany}
\author{A.A. Katanin}
%\email{a.katanin@fkf.mpg.de}
\affiliation{Max-Planck-Institute for Solid State Research,
Heisenbergstr.~1, D-70569 Stuttgart, Germany}
\affiliation{Institute of Metal Physics, 620219 Ekaterinburg, Russia}
\author{W. Metzner}
%\email{w.metzner@fkf.mpg.de}
\affiliation{Max-Planck-Institute for Solid State Research,
Heisenbergstr.~1, D-70569 Stuttgart, Germany}

\date{\today}

\begin{abstract}
We extend the Hertz-Millis theory of quantum phase transitions 
in itinerant electron systems to phases with {\em broken} discrete 
symmetry.
Using a set of coupled flow equations derived within the functional 
renormalization group framework, we compute the second order phase 
transition line $T_c(\delta)$, with $\delta$ a non-thermal control 
parameter, near a quantum critical point. 
We analyze the interplay and relative importance of quantum and classical 
fluctuations at different energy scales, and we compare the Ginzburg 
temperature $T_{G}$ to the transition temperature $T_{c}$, the latter 
being associated with a non-Gaussian fixed-point.
\end{abstract}
\pacs{05.10.Cc, 73.43.Nq, 71.27.+a}

\maketitle

\section{Introduction}

Quantum phase transitions in itinerant electron systems continue to ignite 
considerable interest.\cite{sachdev_book,belitz_review05,loehneysen_review06}
These transitions are usually induced by varying a non-thermal control 
parameter such as pressure or chemical potential at zero temperature, 
so that no classical fluctuations occur and the transition is driven 
exclusively by quantum effects. 
In many physical situations, a line of finite temperature second 
order phase transitions in the phase diagram terminates at a quantum 
critical point at $T=0$.
In such cases quantum fluctuations influence the system also at 
finite temperatures altering measurable quantities such as the 
shape of the phase boundary. 
Consequently, in a complete description of the system at finite $T$, 
quantum and thermal fluctuations have to be accounted for simultaneously.

The conventional renormalization group (RG) approach to quantum 
criticality in itinerant electron systems \cite{hertz76,millis93} 
relies on the assumption that it is sensible to integrate out fermionic 
degrees of freedom from the functional integral representation of the 
partition function and then to expand the resulting effective action 
in powers of the order parameter alone.
This approach has been questioned for magnetic phase transitions
associated with spontaneous breaking of continuous spin rotation
invariance, since integrating the fermions results in singular 
interactions of the order parameter field.
\cite{belitz_review05,loehneysen_review06}

In this paper, we focus on quantum phase transitions to phases with 
broken {\em discrete} symmetry.
We analyze quantum and classical fluctuations in the {\em symmetry-broken} 
phase with an Ising-like order parameter near a quantum critical point.
Our calculations are based on a set of coupled flow equations obtained
by approximating the exact flow equations of the one-particle irreducible
version of the functional RG.
Quantum and classical (thermal) fluctuations are treated on equal
footing.
The functional RG has been applied extensively to classical critical 
phenomena,\cite{berges_review02} where it provides a unified description 
of $O(N)$-symmetric scalar models, including two-dimensional systems. 
The classical Ising universality class has been analyzed in 
Refs.~\onlinecite{ballhausen04,delamotte03,kopietz07}.
The functional RG is obviously also a suitable framework for treating 
quantum criticality.\cite{wetterich07} 
In our approach, we can compute the RG flow in any region 
of the phase diagram, including the region governed by non-Gaussian 
critical fluctuations. 
This allows comparison between the true transition line and the Ginzburg 
line, which so far has been used as an estimate of the former.
\cite{loehneysen_review06,millis93}
Specifically, we capture the strong-coupling behavior emergent in 
the vicinity of the transition line as well as the correct classical 
fixed-point for $T_c$, including the anomalous dimension of the 
order parameter field.

Our results may be applied to commensurate
charge-density-waves, Pomeranchuk transitions 
\cite{metzner03,dellanna06,rosch07} 
which spontaneously break the discrete point group lattice symmetry,
and to magnetic transitions with Ising symmetry, provided the
transitions are continuous at low temperatures.

The paper is structured as follows: in Section II we introduce Hertz's
\cite{hertz76} effective action, adapted to the symmetry-broken phase,
which serves as a starting point for the subsequent analysis. 
In Section III we describe the functional
RG method and its application in the present context, and subsequently 
derive the RG flow equations. In Section IV we present a solution of the 
equations in the case $T=0$. Section V contains numerical results for 
the finite $T$ phase diagram in the region with broken symmetry. 
Different cases are discussed, distinguished by the dimensionality $d$ 
and the dynamical exponent $z$. In particular, we compare the $T_c$ 
line with the Ginzburg line, thus providing an estimate of the critical 
region size. In Section VI we summarize and discuss the results.

\section{Hertz action}

The starting point of the standard RG approach to quantum critical 
phenomena in itinerant electron systems is the Hertz action.
\cite{hertz76,millis93}
It can be derived from a microscopic Hamiltonian by applying a 
Hubbard-Stratonovich transformation to the path-integral representation 
of the partition function and subsequently integrating out the fermionic 
degrees of freedom. 
The resulting action is then expanded in powers of the order parameter 
field, usually to quartic order. 
The validity of this expansion is dubious in several physically 
interesting cases, in particular for magnetic transitions with 
SU(2)-symmetry, since the integration over gapless fermionic modes 
can lead to singular effective interactions of the order parameter 
field, which may invalidate the conventional power counting.
\cite{belitz_review05,loehneysen_review06,abanov04}
Such complications probably do not affect transitions in the charge
channel and magnetic transitions with Ising symmetry.
There are several indications that singularities cancel in that case,
namely the cancellation of singularities in effective interactions 
upon symmetrization of fermion loops,\cite{neumayr98,kopper01} and 
the cancellation of non-analyticities in susceptibilities.
\cite{chubukov03,rech06}
We therefore rely on the usual expansion of the action to quartic 
order in the order parameter field and also on the conventional
parametrization of momentum and frequency dependences,
which leads to the Hertz action \cite{hertz76,millis93} 
\begin{eqnarray}
 S[\phi] = 
 \frac{T}{2} \sum_{\omega_n} \int\frac{d^dp}{(2\pi)^d} \,
 \phi_p \left( \frac{|\omega_{n}|}{|\mathbf{p}|^{z-2}}
 + \mathbf{p}^{2} \right) \phi_{-p} + U[\phi] \; .
 \label{eq:lagrangian}
\end{eqnarray}
Here $\phi$ is the scalar order parameter field and $\phi_p$ with
$p = (\mathbf{p},\omega_n)$ its momentum representation;
$\omega_n = 2\pi n T$ with integer $n$ denotes the (bosonic) 
Matsubara frequencies. 
Momentum and energy units are chosen such that the prefactors 
in front of $\frac{|\omega_{n}|}{|\mathbf{p}|^{z-2}}$ and 
$\mathbf{p}^{2}$ are equal to unity. 
The action is regularized in the ultraviolet by
restricting momenta to $|\mathbf{p}| \leq \Lambda_0$. 
The value of the dynamical exponent is restricted to $z \geq 2$. 
The cases $z=3$, applicable to Pomeranchuk instabilities and
Ising ferromagnetism, and $z=2$, relevant for commensurate 
charge density waves and Ising antiferromagnetism, 
are of our main interest. 
Formally, the results of the paper may be applied 
to systems with arbitrary $z\geq 2$. 
A comprehensive discussion of the origin of the $\omega_n$ dependence 
of the action is given by Millis \cite{millis93} (see also 
Ref.~\onlinecite{Nagaosa_book}).
Eq.~(\ref{eq:lagrangian}) is valid in the limit
$\frac{|\omega_{n}|}{|\mathbf{p}|^{z-2} } \ll 1$ which is relevant 
here since the dominant fluctuations occur only in this regime.

In the symmetric phase the potential $U[\phi]$ is minimal at
$\phi=0$, and is usually parametrized by a positive quadratic 
and a positive quartic term.\cite{hertz76,millis93}
Since we approach the quantum critical point from the symmetry-broken 
region of the phase diagram, we assume a potential 
$U[\phi]$ with a minimum at a non-zero order parameter $\phi_0$:
\begin{eqnarray}
 U[\phi] &=& \frac{u}{4!} \int_0^{1/T} d\tau \int d^d x 
 \left( \phi^{2} - \phi_{0}^{2} \right)^{2} \nonumber\\
 &=& \int_0^{1/T} \!\!\! d\tau \int \! d^d x
 \left[u\,\frac{\phi'^{4}}{4!}+\sqrt{3 \, u\, \delta} \,
 \frac{\phi'^{3}}{3!} + \delta \, \frac{\phi'^{2}}{2!}\right] \, ,
\label{eq:ef_potential}
\end{eqnarray}
where $\phi$ and $\phi'$ are functions of $x$ and $\tau$ 
with $\phi= \phi_{0}+\phi'$. 
The parameter $\delta = u\,\phi_{0}^{2}/3$ controls the 
distance from criticality. 
Approaching the phase boundary in the $(\delta, T)$-plane from 
the symmetry-broken phase gives rise to the three-point vertex 
$\sqrt{3\, u\, \delta}$, which generates an anomalous dimension 
of the order parameter field already at one-loop level.

Although formally correct as a result of integrating out the
fermions, the Hertz action (1) is not a good starting point 
for symmetry-broken phases with a fermionic gap, 
such as charge density wave phases or antiferromagnets.
A fermionic gap leads to a suppression of the dynamical 
term (linear in frequency) in the action, since it suppresses
low energy particle-hole excitations.
If not treated by a suitable resummation in the beginning, 
this effect is hidden in high orders of perturbation theory.
\cite{rosch01}
We do not deal with this complication in the present paper.
However, our results for the transition temperature should not 
be affected by a gap in the symmetry-broken phase, since it
vanishes continuously at $T_c$.

\section{Method}

To analyze the quantum field theory defined by $S[\phi]$
we compute the flow of the effective action $\Gamma^{\Lambda}[\phi]$
with approximate flow equations, which are derived from an exact
functional RG flow equation
\cite{wetterich93, berges_review02, metzner_review, delamotte07, gies06}.
The effective action $\Gamma^{\Lambda}[\phi]$ is the generating 
functional for one-particle irreducible vertex functions in presence
of an infrared cutoff $\Lambda$. The latter is implemented by adding a 
regulator term of the form 
$\int \frac{1}{2} \phi R^{\Lambda} \phi$ to the bare action.
The effective action interpolates smoothly between the bare action
$S[\phi]$ for large $\Lambda$ and the full effective action 
$\Gamma[\phi]$ in the limit $\Lambda \to 0$ (cutoff removed).
Its flow is given by the exact functional equation 
\cite{wetterich93}
\begin{eqnarray}
\frac{d}{d \Lambda}\Gamma^{\Lambda}\left[\phi\right]=
\frac{1}{2}\text{Tr}\frac{\dot{R}^{\Lambda}}{\Gamma^{(2)}\left[\phi\right]
+ R^{\Lambda}}\,\,,
\label{eq:flow_eqn}
\end{eqnarray}
where $\dot{R}^{\Lambda}=\partial_{\Lambda}R^{\Lambda}$, and
$\Gamma^{(2)}\left[\phi\right] = \delta^{2}\Gamma^{\Lambda}[\phi]/
 \delta \phi^{2}$.
In momentum representation ($\phi_p$), 
the trace sums over momenta and frequencies:
$\text{Tr} = T \sum_{\omega_{n}} 
 \int \frac{d^{d} p}{\left(2\pi\right)^{d}}$.
For the regulator function $R^{\Lambda}(\mathbf{p})$ we choose the 
optimized Litim cutoff \cite{litim01} 
\begin{equation}
 R^{\Lambda}(\mathbf{p}) = 
 Z \left( \Lambda^{2}-\mathbf{p}^{2}\right)
 \theta\left(\Lambda^{2}-\mathbf{p}^{2} \right) \; ,
\label{Litim_fun}
\end{equation}
where $Z$ is a renormalization factor (see below).

We approximate $\Gamma^{\Lambda}[\phi]$ by a simplified ansatz with a 
local potential of the form Eq. (\ref{eq:ef_potential}) characterized by 
cutoff dependent parameters $u$ and $\phi_0$ (or, alternatively, $\delta$),
and a simple renormalization of the momentum and frequency dependences of
the quadratic part of the action via $Z$-factors, leading to an inverse
propagator of the form
\begin{equation}
G^{-1}(\textbf{p},\omega_n) = \Gamma^{(2)}\left[\phi=\phi_0\right] = 
Z_{\omega} \frac{|\omega_{n}|}{|\mathbf{p}|^{z-2}} + 
 Z\mathbf{p}^2 + \delta + R^{\Lambda}(\textbf{p}) \; ,
\label{Green}
\end{equation}
where the $Z$-factors depend on the scale $\Lambda$ only.
The regulator function $R^{\Lambda}(\textbf{p})$ replaces
$Z\mathbf{p}^2$ with $Z \Lambda^2$ for
$|\mathbf{p}| < \Lambda$.

Evaluating Eq.~(\ref{eq:flow_eqn}) for a momentum-independent field $\phi$ 
yields the flow of the effective potential $U[\phi]$
\begin{equation}
 \partial_\Lambda U[\phi] = 
 \frac{1}{2} \, \text{Tr} \, \frac{\dot{R}^{\Lambda}(\mathbf{p})}
 {Z_{\omega}\frac{|\omega_n|}{\textbf{p}^{z-2}} + 
 Z\textbf{p}^2 + R^\Lambda(\mathbf{p}) + U''[\phi]}\; ,
\label{ef_pot_flow}
\end{equation}
from which we derive the flows of the parameters $\phi_0$ and $u$,
following the procedure in Ref.~\onlinecite{berges_review02}.
Viewing $U$ as a function of $\rho = \frac{1}{2} \phi^2$ and using
$U'[\rho_0] = 0$, we can write
$0 = \frac{d}{d\Lambda} U'[\rho_0] = 
 \partial_\Lambda U'[\rho_0] +
 U''[\rho_0] \, \partial_\Lambda\rho_0$.
Inserting $\partial_\Lambda U'[\rho_0]$ as obtained by differentiating 
Eq.~(\ref{ef_pot_flow}) with respect to $\rho$ at $\rho = \rho_0$,
and using $U''[\rho_0] = \frac{1}{3} u$,
one obtains the flow equation for $\rho_0$
\begin{equation}
 \partial_\Lambda \rho_0 =
 \frac{3}{2} \, \text{Tr} \, \frac{\dot{R}^{\Lambda}(\mathbf{p})}
 {\left[Z_{\omega}\frac{|\omega_n|}{\textbf{p}^{z-2}} + 
 Z\textbf{p}^2 + R^\Lambda(\mathbf{p}) + 
 \frac{2}{3} u \rho_0 \right]^2} \; .
\label{rho0_flow}
\end{equation}
The flow of $u$ is obtained by differentiating Eq.~(\ref{ef_pot_flow}) 
twice with respect to $\rho$:
\begin{equation}
 \partial_\Lambda u =
 3 u^2 \, \text{Tr} \, \frac{\dot{R}^{\Lambda}(\mathbf{p})}
 {\left[Z_{\omega}\frac{|\omega_n|}{\textbf{p}^{z-2}} + 
 Z\textbf{p}^2 + R^\Lambda(\mathbf{p}) + 
 \frac{2}{3} u \rho_0 \right]^3} \; .
\label{u_flow}
\end{equation}
Inserting the above flow equations into the $\Lambda$-derivative
of $\delta = \frac{2}{3} u \rho_0$, we obtain the flow of 
$\delta$.
Below we will analyze the flow in terms of $u$ and $\delta$ 
instead of $u$ and $\rho_0$.

The Matsubara frequency sums in the flow equations (\ref{rho0_flow}) 
and (\ref{u_flow}) can be expressed in terms of polygamma functions
$\Psi_n(z)$, defined recursively by 
$\Psi_{n+1}(z) = \Psi_n'(z)$ for $n=0,1,2,\dots$, and
$\Psi_0(z) = \Gamma'(z)/\Gamma(z)$, where $\Gamma(z)$ is the
gamma function. 
From the Weierstrass representation of the gamma function,
$\Gamma(z)^{-1} = 
 z e^{\gamma z}\prod_{n=1}^{\infty}(1+\frac{z}{n})e^{-z/n}$,
where $\gamma$ is the Euler constant, one can derive
the relation
\begin{equation}
 \sum_{n=-\infty}^{\infty} \frac{1}{(|n|+z)^2} =
 \frac{1}{z^2} + 2 \Psi_1(z+1) \; .
\label{freqsum}
\end{equation}
Taking derivatives with respect to $z$ yields expressions for
sums involving higher negative powers of $(|n|+z)$ in terms of 
polygamma functions of higher order.

For $R^{\Lambda}(\mathbf{p})$ we now insert the Litim function 
defined in Eq. (\ref{Litim_fun}).
Following Ref.~\onlinecite{berges_review02} we neglect 
$\dot{Z}$ 
in $\dot{R}^{\Lambda}(\mathbf{p})$, such that
$\dot{R}^{\Lambda}(\mathbf{p}) = 
 2 Z \Lambda \Theta(\Lambda^2 - \mathbf{p}^2) \,$.
The neglected terms would yield corrections to the flow equations that are linear in $\eta$ and therefore irrelevant in the quantum part of the flow, where $\eta\approx 0$ (see below). In the classical part of the flow (for $\Lambda\to 0$) these corrections are small as compared to other included terms involving $\eta$  due to the presence of additional factors involving the interaction coupling and mass. Therefore they give rise to only minor corrections to the values of the fixed point coordinates and the values of the critical exponents.

The $d$-dimensional momentum integrals on the right hand side of
the flow equations can be reduced to one-dimensional integrals,
since the integrands depend only on the modulus of $\mathbf{p}$.

Explicit dependences on $\Lambda$, $Z$-factors, and lengthy
numerical prefactors in the flow equations can be eliminated by 
using the following rescaled variables:
\begin{eqnarray}
 \tilde p &=& |\mathbf{p}|/\Lambda \; , \\
 \tilde T &=& 
 \frac{2\pi Z_{\omega}}{Z\Lambda^z} \, T \; , \\
 \tilde\delta &=& \frac{\delta}{Z\Lambda^2} \; , \\
 \tilde u &=& 
 \frac{A_d T}{2 Z^2 \Lambda^{4-d}} \, u \; ,
\end{eqnarray}
with $A_d = (2\pi)^{-d} S_{d-1}$, where
$S_{d-1} = 2\pi^{d/2}/\Gamma(d/2)$ is the area of the 
$(d\!-\!1)$-dimensional unit sphere.

The flow equations for $\tilde\delta$ and $\tilde u$ are then 
obtained as
\begin{eqnarray}
 \frac{d\tilde\delta}{d\log\Lambda} &=& 
 (\eta - 2) \, \tilde\delta  \nonumber \\
 &+& \!
 4 \tilde{u} \, \Bigg[ 
 \frac{1}{d} \frac{1}{(1+\tilde{\delta})^2} +
 \frac{2}{\tilde T^2} \int_0^1 \! d \tilde p \,
 \tilde p^{d+2z-5} \Psi_1(x) \Bigg]  \nonumber \\
 &+& \!
 12 \tilde u \tilde\delta \, \Bigg[
 \frac{1}{d} \frac{1}{(1+\tilde\delta)^3} - 
 \frac{1}{\tilde T^3} \int_0^1 \! d\tilde p \,
 \tilde p^{d+3z-7} \Psi_2(x) \Bigg] \; , \hskip 7mm
\label{eq:delta_delta}
\end{eqnarray}
\begin{eqnarray}
 \frac{d\tilde u}{d\log\Lambda}
  &=& 
 (d-4+2\eta) \, \tilde{u}  \nonumber\\
 &+& \!
 12 \tilde u^2 \Bigg[
 \frac{1}{d} \frac{1}{(1+\tilde{\delta})^3} -
 \frac{1}{\tilde T^3} \int_0^1 \! d\tilde p \, 
 \tilde p^{d+3z-7} \Psi_2(x) \Bigg] \; , \hskip 7mm
\label{eq:delta_u}
\end{eqnarray}
where $x = 1 + \tilde T^{-1} (1 + \tilde\delta) \tilde p^{z-2}$,
and $\eta$ is the anomalous dimension defined as
\begin{equation}
 \eta = - \frac{d\log Z}{d\log\Lambda} \; .
\label{eta}
\end{equation}

To complete the system of flow equations one still needs to derive the 
evolution of $Z$ and $Z_{\omega}$, which parametrize the
momentum and frequency dependence of the propagator. 
Taking the second functional derivative of Eq.~(\ref{eq:flow_eqn}), we
obtain the flow equation for the propagator
\begin{equation}
 \partial_\Lambda G^{-1}(p) = 
 3u \delta \, \text{Tr} \left[ \dot{R}^{\Lambda}(\mathbf{q}) \,
 G^2(q) \, G(q+p) \right],
\label{Gamma2flow}
\end{equation}
where $G(q)$ is given by Eq.~(\ref{Green}).
Here we skipped the contribution from the tadpole diagram, 
since it involves no dependence on momentum and frequency, and therefore 
does not contribute to the flow of the $Z$-factors.
\begin{figure}%[ht]
\begin{fmffile}{20070802_loops_1}
\begin{eqnarray}
\tilde\delta &:&
\parbox{25mm}{\unitlength=1mm\fmfframe(2,2)(1,1){
\begin{fmfgraph*}(15,20)\fmfpen{thin} %pp-with rpa with ladder,  inset
 \fmfleft{l1}
 \fmfright{r1}
 \fmftop{v1}
 \fmfpolyn{full,tension=0.6}{G}{4}
 \fmf{dbl_wiggly,straight}{l1,G4}
 \fmf{dbl_wiggly,straight}{G1,r1}
 \fmffreeze
\fmf{dbl_wiggly,tension=0.1,right=0.7}{G2,v1}
\fmf{dbl_wiggly,tension=0.1,right=0.7}{v1,G3}
\end{fmfgraph*}
}}
+
\parbox{25mm}{\unitlength=1mm\fmfframe(2,2)(1,1){
\begin{fmfgraph*}(20,18)\fmfpen{thin} %pp-with rpa with ladder,  inset
 \fmfleft{l1}
 \fmfright{r1}
 \fmfpolyn{full,tension=0.3}{G}{3}
 \fmfpolyn{full,tension=0.3}{K}{3}
  \fmf{dbl_wiggly}{l1,G1}
 \fmf{dbl_wiggly,tension=0.2,right=0.8}{G2,K3}
 \fmf{dbl_wiggly,tension=0.2,right=0.8}{K2,G3}
 \fmf{dbl_wiggly}{K1,r1}
 \end{fmfgraph*}
}}\nonumber\\[-11mm]
\tilde{u} &:&
\parbox{20mm}{\unitlength=1mm\fmfframe(2,2)(1,1){
\begin{fmfgraph*}(20,12)
\fmfpen{thin} %pp-with rpa with ladder,  inset
\fmfleftn{l}{2}\fmfrightn{r}{2}
\fmfrpolyn{full,tension=0.9}{G}{4}
\fmfpolyn{full,tension=0.9}{K}{4}
\fmf{dbl_wiggly}{l1,G1}\fmf{dbl_wiggly}{l2,G2}
\fmf{dbl_wiggly}{K1,r1}\fmf{dbl_wiggly}{K2,r2}
\fmf{dbl_wiggly,left=.5,tension=.2}{G3,K3}
\fmf{dbl_wiggly,right=.5,tension=.2}{G4,K4}
\end{fmfgraph*}
}}
\nonumber\\[-7mm]
\eta &:&
\parbox{20mm}{\unitlength=1mm\fmfframe(2,2)(1,1){
\begin{fmfgraph*}(20,18)\fmfpen{thin} %pp-with rpa with ladder,  inset
 \fmfleft{l1}
 \fmfright{r1}
 \fmfpolyn{full,tension=0.3}{G}{3}
 \fmfpolyn{full,tension=0.3}{K}{3}
  \fmf{dbl_wiggly}{l1,G1}
 \fmf{dbl_wiggly,tension=0.2,right=0.8}{G2,K3}
 \fmf{dbl_wiggly,tension=0.2,right=0.8}{K2,G3}
 \fmf{dbl_wiggly}{K1,r1}
 \end{fmfgraph*}
}}\nonumber\\[-15mm]\nonumber
\end{eqnarray}
\end{fmffile}
\caption{Feynman diagrams representing the contributions to the flow 
 equations (\ref{eq:delta_delta},\ref{eq:delta_u},\ref{eq:etas}).}
\label{fig:fRG_flow}
\end{figure}
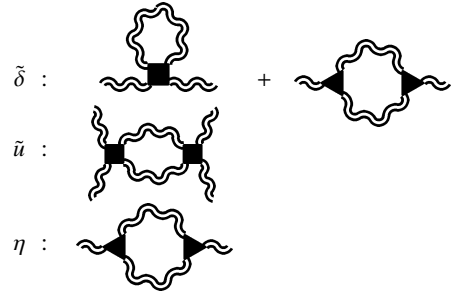

The momentum renormalization factor is given by
\begin{equation}
 Z = \left.
 \frac{1}{2d} \, \Delta_{\textbf{p}}
 \left[G^{-1}(\textbf{p},\omega_n=0) \right] 
 \right|_{\textbf{p} = 0} \; ,
\label{Zp}
\end{equation}
where $\Delta_{\textbf{p}}$ is the Laplace operator evaluated at 
constant cutoff function, that is, $\Delta_{\textbf{p}}$ does not
act on $R^{\Lambda}(\mathbf{p})$. 
Inserting Eqs.~(\ref{Gamma2flow}) and (\ref{Zp}) into Eq.~(\ref{eta}), 
and performing the frequency sum,
one obtains the flow equation for the anomalous dimension
\begin{eqnarray}
 \eta &=&
 \frac{6}{d}\frac{\tilde{u}\tilde{\delta}}{(1+\tilde{\delta})^5}
 \Big[2(1+\tilde{\delta})-\frac{8}{d+2}\Big] - 
 \frac{\tilde{u}\tilde{\delta}}{d} {\tilde T}^{-5}
 \int_0^1 d{\tilde p} {\tilde p}^{d+3z-13}
\nonumber\\
&&
 \Bigg[-6 {\tilde p}^4 (z-2)(d+z-4) \tilde{T}^2
 \Psi_2(x) - 2{\tilde p}^{2+z} \Big(2(8-d)
\nonumber\\
&&
 (1 + \tilde{\delta} - {\tilde p}^2) +
 [(d-14)(1+ \tilde\delta) + 8{\tilde p}^2]z + 3(1+\tilde\delta) z^2 \Big)
\nonumber\\
&&
 \tilde{T} \Psi_3(x) - {\tilde p}^{2z}
 [2({\tilde p}^2 - 1) + \tilde\delta(z-2)+z]^2 \Psi_4(x) \Bigg] \; .
\label{eq:etas}
\end{eqnarray}
The contributions to the flow of $\tilde\delta$, $\tilde u$ and $\eta$ 
are illustrated in terms of Feynman diagrams in Fig.~\ref{fig:fRG_flow}.

We now turn to the renormalization of $Z_{\omega}$.
We first consider the case $z=2$, where $Z_{\omega}$ can be related 
to the propagator by
\begin{equation}
 Z_{\omega} = \frac{1}{2\pi T} \left[
 G^{-1}(\textbf{p}=0,2\pi T) -  G^{-1}(\textbf{p}=0,0) \right]   
 \; .
\end{equation}
Taking the logarithmic derivative with respect to $\Lambda$, inserting
Eq.~(\ref{Gamma2flow}), and performing the trace yields
\begin{equation}
 \eta_{\omega} = 
 \frac{12}{d} \,\tilde{u} \tilde\delta \tilde{T}^{-1}
 \left[\frac{\tilde{T} - (1+\tilde\delta)}{\tilde{T} (1+\tilde\delta)^3} +
 \tilde{T}^{-3} \Psi_2 \left(1 + \frac{1+\tilde\delta}{\tilde{T}} \right)
 \right] \; .
\end{equation}
From the numerical solution of the flow equation we observe that $\eta_\omega$
is small at all scales and vanishes for $\Lambda \to 0$.
For example, in two dimensions $\eta_\omega$ varies between $-0.033$ and 
$0.005$ for $u=1$ at $T=e^{-4}$ and has practically no influence on the 
phase diagram. 
In three dimensions the values are at least one order of magnitude lower. 
For $z>2$ one expects an even smaller $\eta_{\omega}$, since a larger $z$ 
reduces the strength of fluctuations near the quantum critical point.
Therefore we set $\eta_\omega = 0$ and $Z_{\omega}=1$ from now on.
The scaling variable $\tilde{T}$ then obeys the flow equation
\begin{equation}
 \frac{d \tilde T}{d\log\Lambda} =
 (\eta - z) \tilde T \; .
\end{equation}

In the flow equations (\ref{eq:delta_delta},\ref{eq:delta_u},\ref{eq:etas}) 
one identifies the classical mean-field (involving only one power of 
$\tilde{\delta}$ or $\tilde{u}$), classical non-gaussian and quantum 
(involving $\tilde T$) terms. 
The quantum contributions vanish as the infrared cutoff tends to zero at 
constant non-zero temperature ($\tilde{T}^{-1} \ll 1$). 
On the other hand the quantum terms dominate the 
high energy part of the flow, where $\tilde{T}^{-1} \gg 1$.
Our  framework allows for a continuous connection of these two regimes 
of the flow.
The scale $\Lambda_{cl}$ below which quantum fluctuations become irrelevant 
depends on temperature. It vanishes at $T=0$. In the absence of a 
sizable anomalous dimension $\eta$ in the quantum regime of the flow, 
one has
\begin{equation}
 \Lambda_{cl} \propto T^{1/z} \; ,
\label{Lambda_cl}
\end{equation}
as follows directly from the definition of $\tilde T$. It turns out
that $\eta$ is indeed negligible down to the scale $\Lambda_{cl}$, 
except in the case $z=2$ in two dimensions (see Sec. \ref{sec:finite}).

We emphasize that we have used a relatively simple parametrization 
of the effective action $\Gamma^{\Lambda}[\phi]$. In particular, we kept only the 
dominant term in the derivative expansion \cite{berges_review02} and 
neglected the field dependence of the $Z$-factors. 
Furthermore, the simple parametrization of the effective potential 
Eq.~(\ref{eq:ef_potential}) allowed us to substitute the partial 
differential equation (\ref{ef_pot_flow}) governing the flow of $U[\phi]$ 
by the two ordinary differential equations (\ref{eq:delta_delta}) and 
(\ref{eq:delta_u}). 
The latter approximation is equivalent to neglecting all higher order
vertices generated during the flow.  More sophisticated truncations have
been applied in the context of the classical Ising universality class, 
where they lead to improved results for the critical
exponents.\cite{ballhausen04, delamotte03}

\section{Solution at zero temperature}

In this section we present a solution of the flow equations 
(\ref{eq:delta_delta},\ref{eq:delta_u},\ref{eq:etas}) in the case 
of zero temperature, recovering earlier results for the critical 
exponents.
%%% Ref.?
We also provide an analytic expression for the value of the control 
parameter $\delta_0$ corresponding to the quantum critical point.

For this purpose we first take advantage of the asymptotic properties 
of the polygamma functions $\Psi_1(x) = x^{-1} + \mathcal{O}(x^{-2})$ 
for $x \gg 1$, $\Psi_{n+1}(x) = \Psi_n'(x)$, and evaluate the integrals 
in Eqs.~(\ref{eq:delta_delta},\ref{eq:delta_u},\ref{eq:etas}) in the 
regime $\tilde{T} \ll 1$. 
We introduce the variable
\begin{equation}
 \tilde v = \frac{\tilde u}{\tilde T} =
 \frac{A_d}{4\pi Z_{\mathbf{p}} Z_{\omega} \, \Lambda^{4-d-z}} \, u
\end{equation}
and then pass to the limit $T\to 0$. Note that $\tilde v$ does not depend
on temperature.
The resulting flow equations are
\begin{eqnarray}
 \frac{d\tilde\delta}{d\log\Lambda} &=&
 (\eta-2)\tilde{\delta} + 
 \frac{12\tilde{\delta} \tilde v}{(1+\tilde\delta)^2}\frac{1}{d+z-2} + 
 \frac{8 \tilde v}{1+\tilde\delta}\frac{1}{d+z-2} \; , \nonumber \\
 \frac{d\tilde v}{d\log\Lambda} &=& (d+z-4+2\eta) \tilde v + 
 \frac{12 \tilde v^2}{(1+\tilde\delta)^2} \frac{1}{d+z-2} \; , 
 \nonumber\\
 \eta &=&
 \frac{6}{d}\frac{\tilde{\delta} \tilde v}{(1+\tilde\delta)^4}
 \left[\frac{4}{3}(1+\tilde\delta) \right. \nonumber \\
 &-& \left. \frac{(d-2)(z-2)}{3(d+z-4)}(1 + \tilde\delta)^2 - 
 \frac{4}{d+z}\right] \; .
\label{eq:beta_f_T0}
\end{eqnarray}
The above expression for $\eta$ is valid for $d+z>4$. 
The flow equations for the case $d=z=2$ are obtained by skipping 
the term involving $(d-2)(z-2)/(d+z-4)$ in Eq.~(\ref{eq:beta_f_T0}). 
Non-gaussian contributions to the above equations originate from the 
quantum part of Eqs.~(\ref{eq:delta_delta},\ref{eq:delta_u},\ref{eq:etas}). 
The classical non-gaussian terms do not survive the limit $T \to 0$.

Eqs.~(\ref{eq:beta_f_T0}) have an infrared stable Gaussian 
fixed point in $\tilde v = 0$, $\tilde\delta = 0$, $\eta = 0$.
Linearizing the flow equations around the fixed point, one obtains
the solution
\begin{eqnarray}
 \tilde\delta(\Lambda) &=& \left[ \tilde\delta^* + 
 \frac{8 \tilde v^*}{(d+z-2)^2}
 \left(\left(\Lambda/\Lambda_0\right)^{d+z-2} - 1 \right) \right] 
 \left(\Lambda/\Lambda_0\right)^{-2} \; , \nonumber \\
 \tilde v(\Lambda) &=& 
 \left(\Lambda/\Lambda_0\right)^{d+z-4} \tilde v^* \; , \nonumber \\[2mm]
 \eta(\Lambda) &=& 0
\label{gdl2}
\end{eqnarray}
for $d+z>4$.
Here $\tilde\delta^*$ and $\tilde v^*$ on the right hand sides denote the 
initial values of the parameters at $\Lambda=\Lambda_0$.
In the marginal case $d+z=4$ one finds logarithmic convergence of 
$\tilde v(\Lambda)$ and $\tilde\delta(\Lambda)$ to zero.
Expressing the order parameter $\phi_0$ in terms of $\tilde\delta^*$ 
and $\tilde v^*$, substituting the above solution, and taking the limit 
$\Lambda \to 0$ yields
\begin{equation}
 \phi_0 \propto \sqrt{\delta - \delta_0} \; ,
\label{gdl}
\end{equation}
where
\begin{equation}
 \delta_0 = \frac{2A_d}{\pi} \, 
 \frac{\Lambda_0^{d+z-2}}{(d+z-2)^2} \, u
\end{equation}
is the quantum critical point's coordinate. From Eq.~(\ref{gdl}) we read 
off the value of the exponent $\beta = \frac{1}{2}$, consistent with 
mean-field theory. From Eq.~(\ref{gdl2}) we can also straightforwardly 
evaluate the correlation length $\xi$, using
$\xi^{-2} = \lim_{\Lambda\to 0} \delta(\Lambda) =
\lim_{\Lambda \to 0} Z \Lambda^2 \tilde{\delta}$, 
which yields
\begin{equation}
 \xi = (\delta - \delta_0)^{-1/2} \; , 
\end{equation}
as expected within mean field theory.

As anticipated,\cite{hertz76,millis93} the quantum phase transitions 
studied here behave similarly to classical transitions with 
effective dimensionality $\mathcal{D} = d+z$.
For $d \geq 2$ and $z \geq 2$ one has $\mathcal{D} \geq 4 $,
leading to mean-field behavior governed by a Gaussian fixed point.\cite{hertz76}
In the next section we turn to finite temperatures, where non-Gaussian
behavior occurs sufficiently close to the phase transition line.

\section{Finite temperatures}
\label{sec:finite}

In this section we numerically solve the RG flow equations 
(\ref{eq:delta_delta},\ref{eq:delta_u},\ref{eq:etas}) for finite 
temperatures. As already announced, the analysis is focused on the
region of the phase diagram where symmetry-breaking occurs. 
In our notation this corresponds to sufficiently large values of the 
control parameter $\delta$. 
First we treat the case $z=3$ (in $d=2,3$) to which 
Eqs.~(\ref{eq:delta_delta},\ref{eq:delta_u},\ref{eq:etas}) are applied 
directly. The case $z=2$, in which additional simplifications occur, 
is analyzed in Subsection B.
In all numerical results we choose an initial cutoff $\Lambda_0 = 1$,
and an initial coupling constant $u = 1$.

\subsection{z = 3}

We first solve the coupled flow equations 
(\ref{eq:delta_delta},\ref{eq:delta_u},\ref{eq:etas}) 
with the aim of determining the phase boundary $T_c(\delta)$, or, 
equivalently $\delta_c(T)$.
To this end, for each given temperature we tune the initial value 
of $\delta$ such that at the end of the flow (for $\Lambda \to 0$) 
one obtains the critical state with $\delta(\Lambda) \to 0$. The tuned
initial value is then identified as $\delta_c(T)$. Inverting the
function $\delta_c(T)$ yields $T_c(\delta)$.
In the variables $\tilde\delta$, $\tilde{u}$ this corresponds to 
seeking for such values of the initial $\delta$, 
that both $\tilde\delta(\Lambda)$ and $\tilde{u}(\Lambda)$ reach a 
fixed point as the cutoff is removed.

\begin{figure}[ht!]
\begin{center}
\includegraphics[width=3.2in]{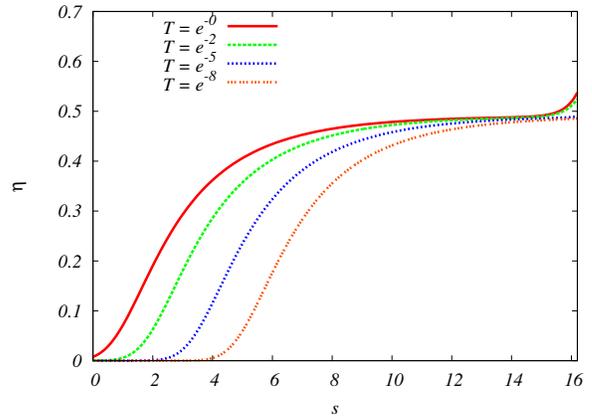}
\caption{(Color online) Anomalous dimension $\eta(\Lambda)$ plotted as function of 
 $s = -\log(\Lambda/\Lambda_0)$ for different values of temperature 
 in the case $z=3$ and $d=2$. The function $\eta(s)$ exhibits crossover
 from the mean-field value $\eta=0$, to the non-gaussian result 
 $\eta \approx 0.48$. 
 The crossover scale $\Lambda_G$ is shifted towards smaller values of $\Lambda$
 (larger $s$) as temperature is reduced.}
\label{fig:eta_vs_lambda}
\end{center}
\end{figure}

\begin{figure}[ht!]
\begin{center}
\includegraphics[width=3.2in]{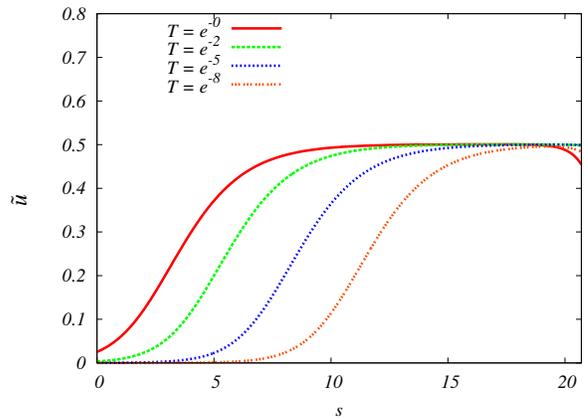}
\caption{(Color online) Quartic coupling $\tilde{u}(\Lambda)$ plotted as function 
 of $s = - \log(\Lambda/\Lambda_0)$ for different values of 
 temperature in the case $z=3$ and $d=3$.}
\label{fig:u_vs_lambda}
\end{center}
\end{figure}

The flow of $\eta$ and $\tilde u$ as a function of the logarithmic
scale variable $s = - \log(\Lambda/\Lambda_0)$ is shown in two
exemplary plots in Figs.~\ref{fig:eta_vs_lambda} and 
\ref{fig:u_vs_lambda}, respectively.
The flow is shown for various temperatures $T$, with $\delta$ tuned 
to values very close to $\delta_c(T)$.
The plateaus in Figs.~\ref{fig:eta_vs_lambda} and \ref{fig:u_vs_lambda} 
correspond to non-gaussian fixed point values of the flowing parameters. 
These values are not altered by the quantum terms in the flow equations.
From Fig.~\ref{fig:eta_vs_lambda}, where $d=2$, one reads off the value of the 
anomalous dimension $\eta\approx 0.48$.
The exact value from the Onsager solution to the Ising model is 
$\frac{1}{4}$. 
For the case $d=3$ we find $\eta \approx 0.08$ within our truncation,
which is also about twice as large as accurate estimates of the
exact value.\cite{Goldenfeld_book}  
These results can also be compared to earlier classical $\phi^4$ RG studies using similar though more sophisticated truncations. In Ref.~\onlinecite{ballhausen04} a derivative expansion to order $\partial^2$, retaining the general form of $U[\phi]$ and field dependence of wave function renormalization, was applied in dimensionality $1<d<4$ yielding $\eta\approx 0.08$ in $d=3$ and $\eta\approx 0.4 $ in $d=2$. In Ref.~\onlinecite{delamotte03} a derivative expansion to order $\partial^4$ yielded $\eta\approx0.033$ in $d=3$.
To obtain scaling behavior in the range of a few orders of magnitude
one needs to fine-tune the initial conditions with an accuracy of
around 15 digits. The breakdown of scaling behavior observed for 
very small $\Lambda$ is due to insufficient accuracy of the initial
value of $\delta$ and numerical errors. 
The plateaus are more extended as we go on fine-tuning the initial
condition. Only exactly {\em at} the critical point true scale 
invariance manifested by plateaus of infinite size is expected. 

The figures also reveal the Ginzburg scale $\Lambda_G$ at which 
non-gaussian fluctuations become dominant, such that the exponent 
$\eta$ attains a non-zero value. 
By fitting a power-law we observe $\Lambda_G \propto T_{c}$ for $d=3$ 
and $\Lambda_G \propto \sqrt{T_{c}}$ for $d=2$ with non-universal 
proportionality factors. 
As expected, $\Lambda_G$ vanishes at the quantum critical point, 
because at $T=0$ the effective dimensionality $\mathcal{D}=d+z$ 
is above the upper critical dimension $d_c = 4$.
At finite temperatures, $\Lambda_G$ is the scale at which $\tilde u$
is promoted from initially small values (of order $T$) to values of 
order one. From the linearized flow equations one obtains
\begin{equation}
 \Lambda_G \propto T_c^{\frac{1}{4-d}}
\end{equation}
in agreement with the numerical results for $d=2$ and $d=3$.
Note that $\Lambda_G \ll \Lambda_{cl}$, since 
$\Lambda_{cl} \propto T_c^{1/3}$ for $z=3$, see Eq.~(\ref{Lambda_cl}). 
Hence anomalous scaling (finite $\eta$) is indeed absent in the regime 
where quantum fluctuations contribute,
and non-Gaussian fluctuations appear only in the classical regime.

As already mentioned, the quantum contributions influence the flow 
only at relatively large $\Lambda$ for $T > 0$. 
In particular, they do not alter any fixed-point values. 
However, at the beginning of the flow 
they dominate over the classical part and therefore are crucial 
for a correct computation of the initial value of $\delta$ leading 
to a scaling solution in the infrared limit.

\begin{figure}[ht!]
\begin{center}
\includegraphics[width=3.0in]{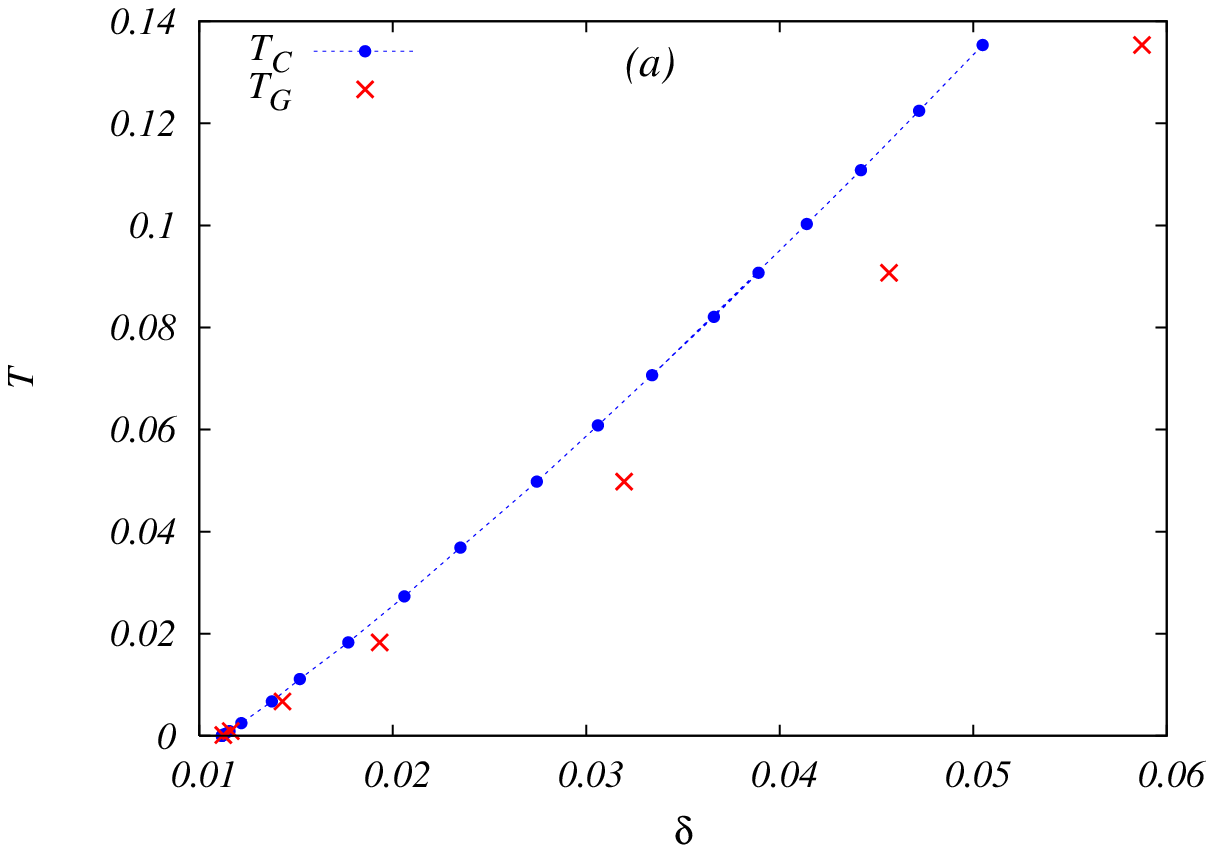}
%\\[10mm]
\includegraphics[width=3.0in]{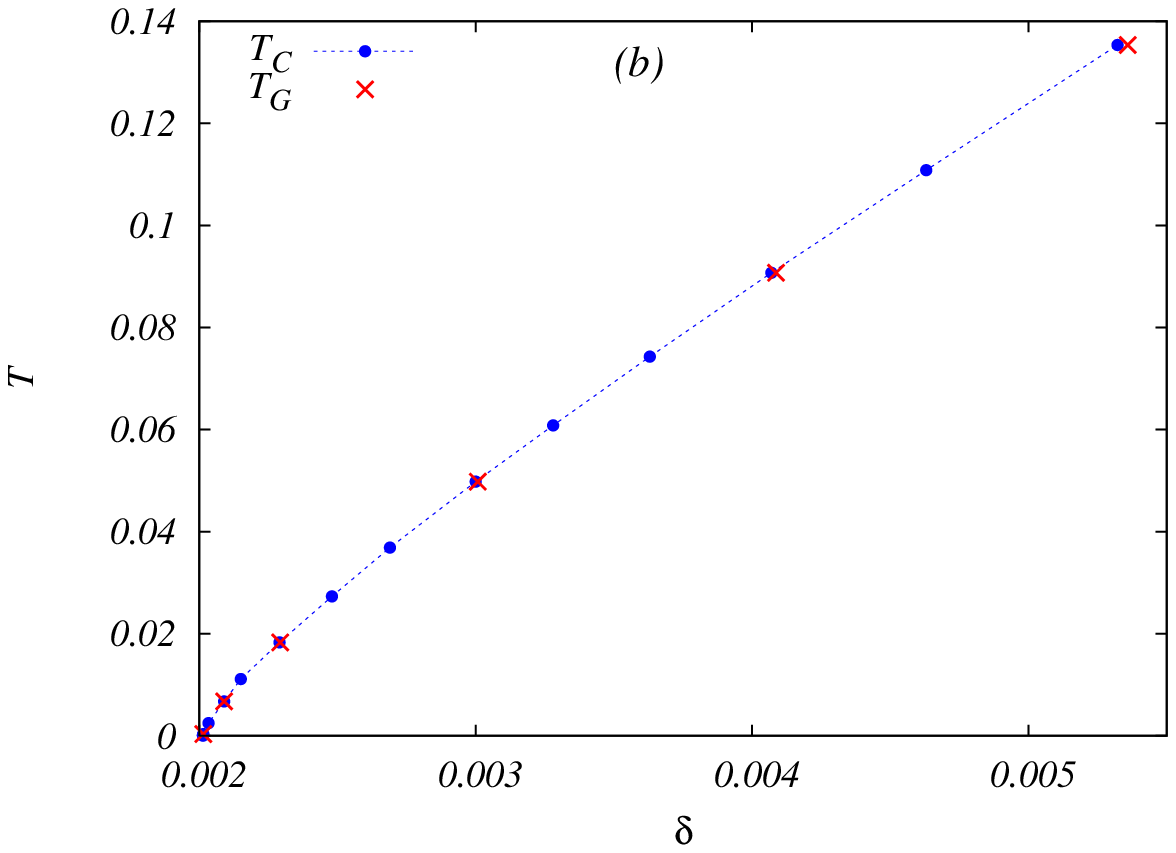}
\caption{(Color online) The transition lines $T_c(\delta)$ obtained for $z=3$,  
 $d=2$ $(a)$ and $z=3$, $d=3$ $(b)$. The phase with broken symmetry is located below the line.
The phase boundary $T_c(\delta)$ obeys 
$(\delta-\delta_0) \propto T_c \log T_c$ for $d=2$ and $T_c \propto (\delta - \delta_0)^{3/4}$ for $d=3$ in agreement with the result 
by Millis.\cite{millis93}
The crosses indicate the Ginzburg temperature $T_G$ for
various choices of $\delta$.}
\label{fig:T_c_vs_delta_d2}
\end{center}
\end{figure}

Results for the transition line $T_c(\delta)$ for $d=2$ and $d=3$ are shown in 
Fig.~\ref{fig:T_c_vs_delta_d2}.
In both cases we recover the shape of the transition line as 
derived by Millis,\cite{millis93} who used the Ginzburg 
temperature in the symmetric phase as an estimate for $T_c$. 
Namely, we find 
\begin{equation}
 (\delta - \delta_0) \propto T_c \log T_c
\end{equation}
for $d = 2$,\cite{comment} and
\begin{equation}
 T_c \propto (\delta - \delta_0)^{3/4}
\end{equation}
for $d = 3$. The exponent $3/4$ in the three-dimensional case 
matches with the general formula for the shift exponent,
$\psi = \frac{z}{d+z-2}$ for arbitrary $z$ in dimensions 
$d > 2$.\cite{millis93}
Note that the phase boundary $T_c(\delta)$ approaches 
the quantum critical point with vanishing first derivative for $d=2$ 
and with singular first derivative in the case $d=3$. In Ref.~\onlinecite{braun06} functional RG methods in a different truncation were applied to calculate the shift exponent in the context of chiral symmetry breaking with the control parameter being the number of fermion flavours.
 
\begin{figure}[ht!]
\begin{center}
\includegraphics[width=3.0in]{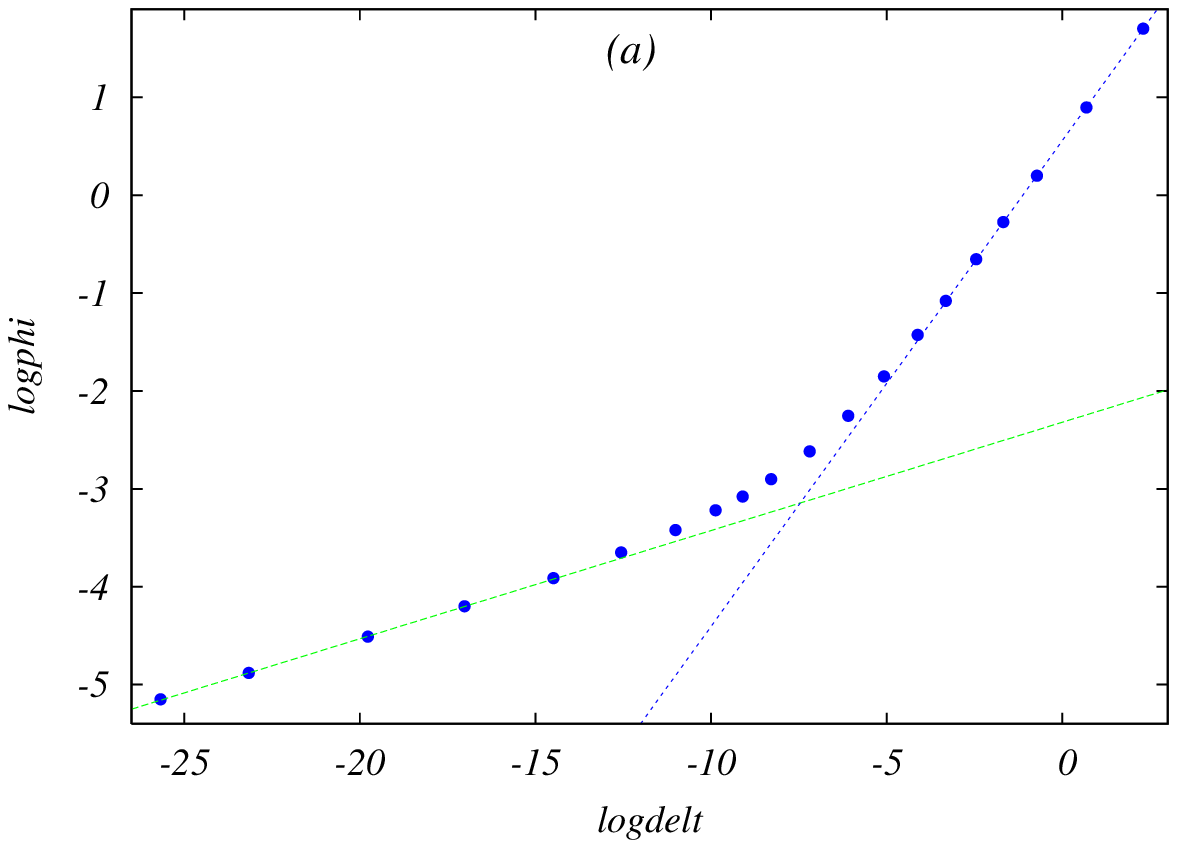}
\includegraphics[width=3.0in]{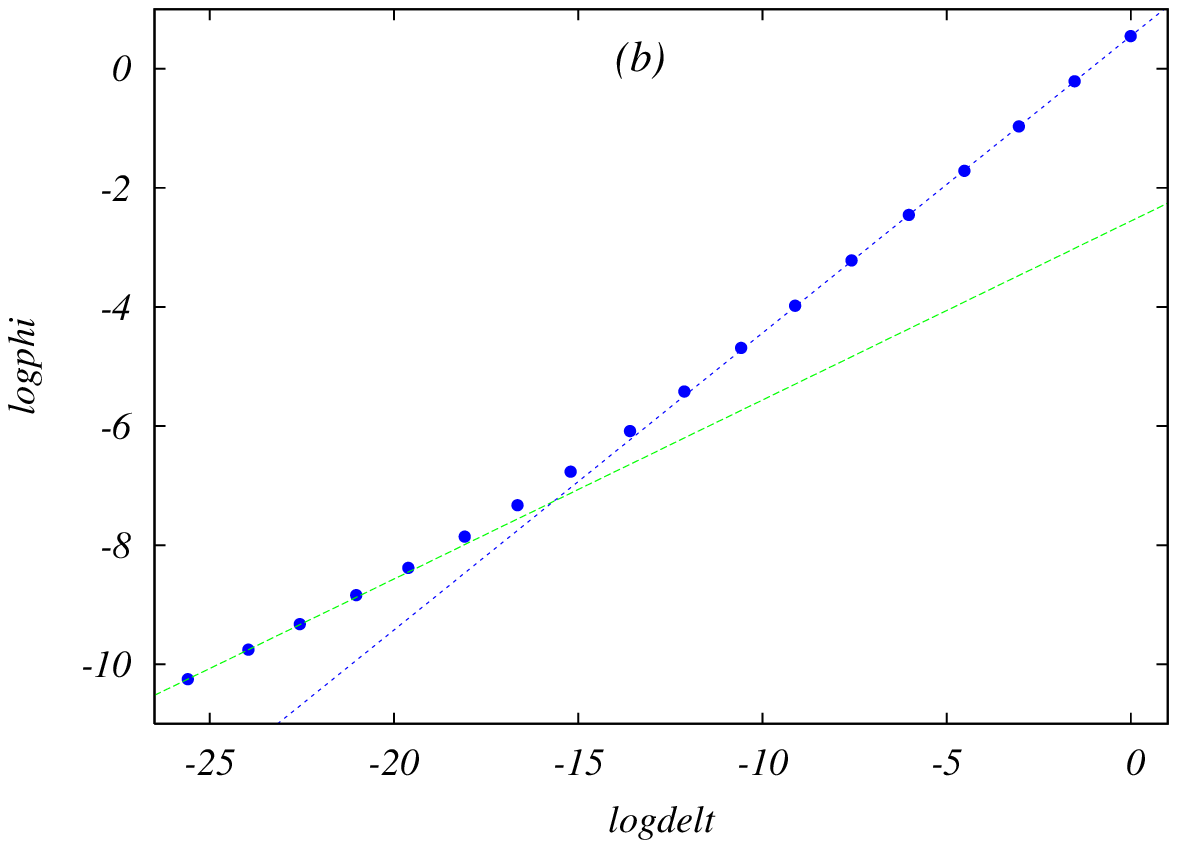}
\caption{(Color online) Order parameter $\phi_0$ as a function of 
$(\delta-\delta_c)$ at $T=e^{-5}$ for $z=3$, $d=2$ $(a)$, 
and $z=3$, $d=3$ $(b)$.
The exponent $\beta$ governing the decay of $\phi_0$ upon 
approaching $\delta_c$ exhibits a crossover from a mean-field 
value $\beta = 0.5$ to a non-gaussian $\beta \approx 0.11$ for $d=2$,  
and $\beta \approx 0.30$ for $d=3$.
The intersection of the straight lines determines $\delta_G$.}
\label{fig:phi_vs_delta_d2}
\end{center}
\end{figure}

An advantage of the present approach is that one can also follow 
the RG flow into the strong coupling regime, where non-gaussian 
critical behavior occurs. This in turn allows an estimate of the 
critical region's size as a function of temperature or the control 
parameter $\delta$. To evaluate the Ginzburg line in the symmetry-broken phase 
one solves the 
flow equations 
(\ref{eq:delta_delta},\ref{eq:delta_u},\ref{eq:etas}) 
for fixed $T$ and at different values of 
$\delta > \delta_c(T)$, observing the behavior of fixed-point 
values of the average order parameter $\phi_0$ (or, alternatively, 
the correlation length $\xi$) as $\delta$ approaches $\delta_c$.
Typical results are plotted in Fig.~\ref{fig:phi_vs_delta_d2} from which 
we read off the value of the exponent $\beta$ describing the 
decay of the order parameter upon approaching the transition 
line $\phi_0 \propto (\delta-\delta_c)^{\beta}$. 
In the truly critical region (for $\delta-\delta_c$ small enough) 
one obtains $\beta\approx 0.11$ for $d=2$ and $\beta\approx 0.30$ 
for $d=3$. These results come out close to the correct classical 
values $0.125$ and $0.31$, respectively. This is unlike the other 
critical exponents ($\eta$ and the correlation length exponent $\nu$) which within 
our truncation differ by factors close to 2 from their correct 
values. Indeed, as discussed in Refs.~\onlinecite{ballhausen04, delamotte03}, 
to obtain accurate values of the critical exponents in the Ising 
universality class, and in particular in $d=2$, one not only needs 
to consider the full partial differential equation governing the RG 
flow of the effective potential $U[\phi]$, but also the field 
dependence of the wave function renormalization and higher orders 
in the derivative expansion of the effective action. 
\cite{berges_review02}

From Fig.~\ref{fig:phi_vs_delta_d2}
we can extract the Ginzburg value $\delta_G$ below which true 
critical behavior is found at the chosen temperature $T$. 
Around $\delta_G$, the exponent $\beta$ exhibits a crossover 
from its mean-field value $\beta = 0.5$ to a non-gaussian value. 
In other words, $\delta_G$ marks the boundary of the non-gaussian 
critical region at a given $T$.
At zero temperature, $\delta_G$ coincides with the quantum
phase transition point $\delta_0$, since there the fluctuations
are effectively $d+z>4$ dimensional, leading to mean-field 
behavior. 
Several Ginzburg points in the $\delta-T$ plane are plotted 
as $T_G(\delta)$ in Fig.~\ref{fig:T_c_vs_delta_d2},
where they can be compared to the phase transition line.
In three dimensions $T_G$ and $T_c$ almost coincide, such that
$T_G$ provides an accurate estimate for $T_c$.
In two dimensions a sizable region between $T_G$ and $T_c$ 
appears in the phase diagram. In that region, non-Gaussian
classical fluctuations are present.

We stress that accounting for the anomalous exponent 
$\eta$ is necessary to describe the classical scaling regime,
that is to obtain the plateaus in Figs.~2,3. 
Upon putting $\eta=0$ the scaling plateaus do not form. 
On the other hand, the shapes of the phase boundaries and 
the Ginzburg curves become very similar at $T\to 0$ in the 
present cases, where $d+z \geq 4$. Therefore our results for 
$T_c(\delta)$ are consistent with earlier studies,\cite{millis93} 
where non-Gaussian critical fluctuations were not captured and the 
calculations were performed in the symmetric phase.

\subsection{z=2}

In the case $z=2$ the flow equations 
(\ref{eq:delta_delta},\ref{eq:delta_u},\ref{eq:etas}) are 
significantly simplified, as all the integrals can be evaluated 
analytically. One obtains
\begin{eqnarray}
 \frac{d\tilde\delta}{d \log\Lambda} 
 &=& \left(\eta - 2\right)\tilde\delta +
 2\tilde{u} \Bigg[ \frac{2}{d}\frac{1}{(1+\tilde\delta)^2} + 
 \frac{4}{d} \tilde{T}^{-2} \Psi_1(y) \Bigg]
 \nonumber \\
 &+&
 3\tilde{u} \tilde\delta
 \Bigg[ \frac{4}{d} \frac{1}{(1+\tilde\delta)^3} - 
 \frac{4}{d} \tilde{T}^{-3}\Psi_2(y) \Bigg] 
 \\[2mm]
 \frac{d\tilde{u}}{d \log\Lambda}
 &=&\left( d-4+2\eta \right) \tilde{u}
 \nonumber \\
 &+&
 3\tilde{u}^2\Bigg[\frac{4}{d}\frac{1}{(1+\tilde\delta)^3}
 - \frac{4}{d} \tilde{T}^{-3} \Psi_2(y) \Bigg] \; ,
\end{eqnarray}
and for the anomalous dimension,
\begin{eqnarray}
 \eta &=&
 \frac{6}{d}\frac{\tilde{u}\tilde\delta}{(1+\tilde\delta)^5}
 \Big[2(1+\tilde\delta)-\frac{8}{d+2}\Big] \nonumber \\
 &+&
 \frac{4\tilde{u} \tilde\delta}{d} \tilde{T}^{-5} 
 \left[\tilde{T} \Psi_3(y)+\frac{1}{d+2}\Psi_4(y)\right] \; ,
\end{eqnarray}
where the argument of the polygamma functions is given by 
$y = 1 + (1+\tilde\delta)/\tilde{T}$.

\begin{figure}[ht!]
\begin{center}
\includegraphics[width=3.0in]{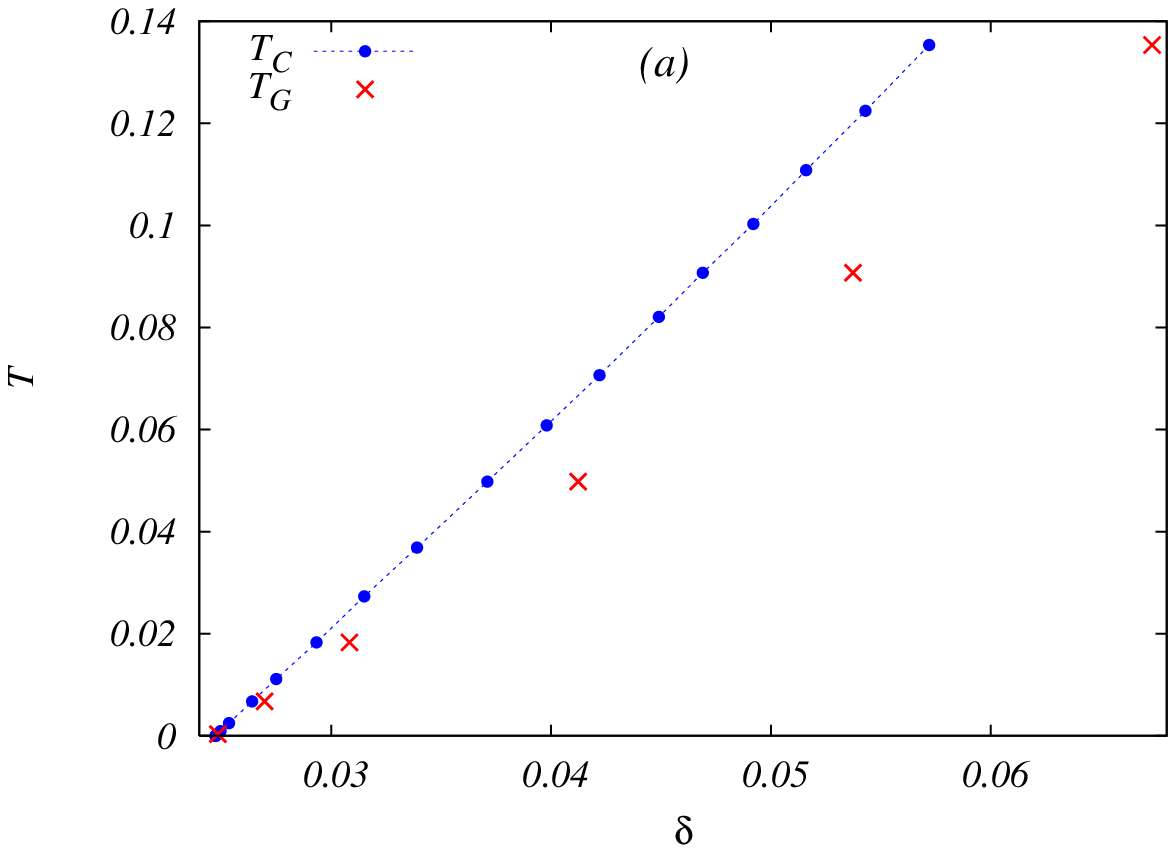}
\includegraphics[width=3.0in]{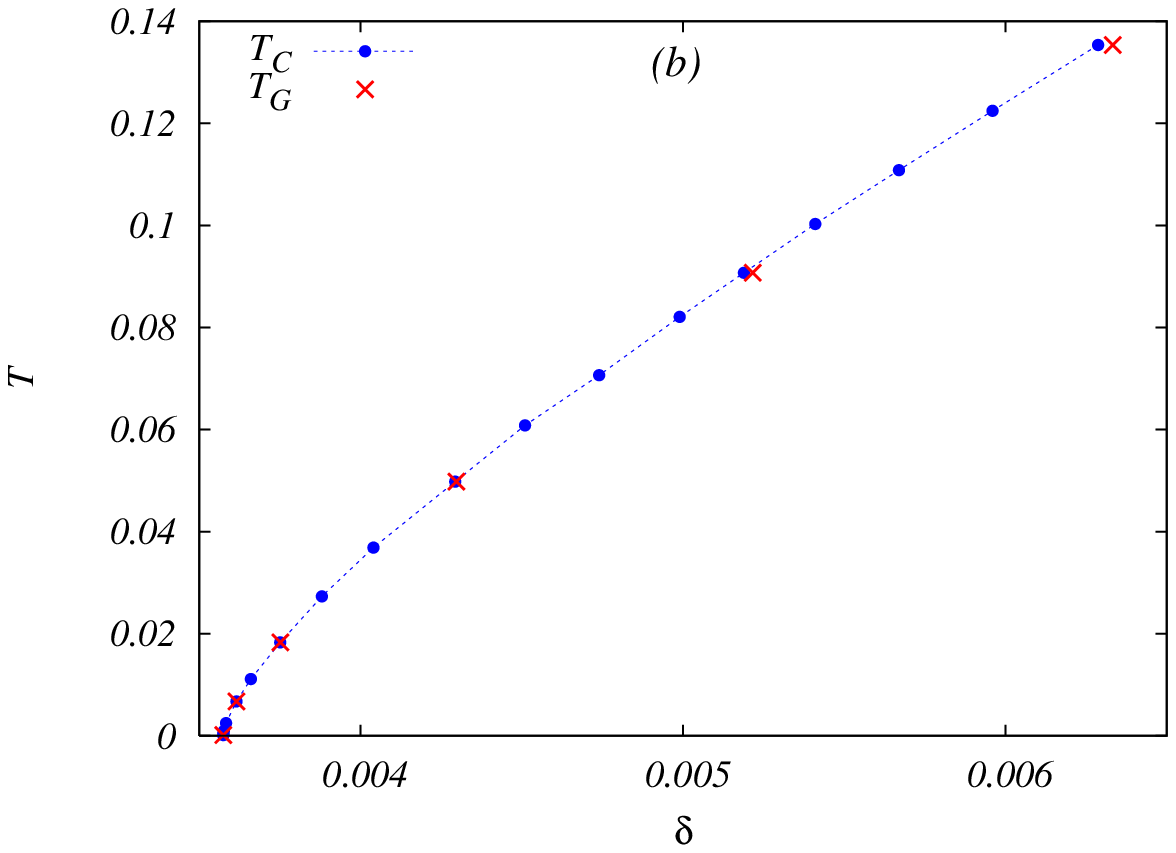}
\caption{(Color online) The transition lines $T_c(\delta)$ obtained for $z=2$,  
 $d=2$ $(a)$ and $z=2$, $d=3$ $(b)$. The phase with broken symmetry is 
 located below the line.
 For $d=2$ the phase boundary $T_c(\delta)$ is consistent with the 
 relation $(\delta-\delta_0) \propto T_c \log\log T_c/\log T_c$ 
 derived by Millis.\cite{millis93}
 For $d=3$ it follows the expected power law 
 $T_c(\delta) \propto (\delta - \delta_0)^{2/3}$. 
 The crosses indicate the Ginzburg temperature $T_G$ for
 various choices of $\delta$.}
\label{fig:T_c_vs_delta_d2z2}
\end{center}
\end{figure}

The procedure to evaluate the phase diagram and the Ginzburg 
line is the same as in the previously discussed case $z=3$. 
In Fig.~\ref{fig:T_c_vs_delta_d2z2} we show results for the transition
line $T_c(\delta)$ in two and three dimensions.
We also show the Ginzburg temperature $T_G$ for various choices 
of $\delta$. 
In two dimensions, the transition line is consistent with the 
almost linear behavior 
$(\delta - \delta_0) \propto T_c \log\log T_c/\log T_c \,$, 
derived previously for the Ginzburg temperature in the symmetric phase.\cite{millis93}
However, a sizable region with non-Gaussian fluctuations 
opens between $T_c$ and $T_G$.
In three dimensions, $T_c(\delta)$ obeys the expected \cite{millis93} 
power law $T_c(\delta) \propto (\delta - \delta_0)^{\psi}$ with
shift exponent $\psi = 2/3$, and the Ginzburg temperature is very
close to $T_c$ for any $\delta$.

\section{Summary}

We have analyzed classical (thermal) and quantum fluctuations in the symmetry-broken 
phase near a quantum phase transition in an itinerant electron
system. The analysis is restricted to the case of {\em discrete} 
symmetry breaking, where no Goldstone modes appear.
Following Hertz\cite{hertz76} and Millis,\cite{millis93} we use an
effective bosonic action for the order parameter fluctuations as a
starting point. 
The renormalization of the Hertz action by fluctuations is obtained from 
a system of coupled flow equations, which are derived as an approximation 
to the exact flow equation for one-particle irreducible vertex functions
in the functional RG framework.
In addition to the renormalization of the effective mass and the four-point 
coupling, we also take the anomalous dimension $\eta$ of the order parameter 
fields into account. In the symmetry-broken phase, contributions to 
$\eta$ appear already at one-loop level.
Quantum and thermal fluctuations are captured on equal footing.
The flow equations are applicable also in the immediate vicinity of the
transition line at finite temperature, where fluctuations deviate strongly
from Gaussian behavior.

We have computed the transition temperature $T_c$ as a function of the
control parameter $\delta$ near the quantum critical point, approaching
the transition line from the symmetry-broken phase. Explicit results
were presented for dynamical exponents $z=2$ and $z=3$ in two and three 
dimensions. In all cases the functional form of $T_c(\delta)$ agrees 
with the behavior of the Ginzburg temperature above $T_c$ derived
previously by Millis.\cite{millis93}
We have also computed the Ginzburg temperature $T_G$ below $T_c$, above 
which non-Gaussian fluctuations become important. While $T_G$ and
$T_c$ almost coincide in three dimensions, a sizable region between
$T_G$ and $T_c$ opens in two dimensions.

It will be interesting to extend the present approach to the case of
continuous symmetry breaking. In that case Goldstone modes suppress
the transition temperature more strongly, and more extended regions
governed by non-Gaussian fluctuations in the phase diagram appear.

\vspace*{1cm}

\begin{acknowledgments}

We are grateful to B.~Delamotte, J.~Pawlowski, A.~Rosch, and 
M.~Salmhofer for valuable discussions, and to J.~Bauer for a 
critical reading of the manuscript.
We also like to thank M.~Kircan for her pleasant collaboration at
the initial stage of the project.
PJ acknowledges support from the German Science Foundation through 
the research group FOR 723 and the Foundation for Polish Science through the START fellowship.
PS acknowledges kind hospitality of G.~Lonzarich and the Quantum 
Matter group at Cavendish Laboratory, University of Cambridge.

\end{acknowledgments}

%%%%%%%%%%%%%%%%%%%%%%%%%%%%%%%%%%%%%%%%%%%%%%%%%%%%%%%%%%%%%%%%%%%%


\begin{thebibliography}{99}
\bibitem{comment} Note that the shape of the Ginzburg curve drawn in 
 Fig.~3 of Ref.~5 does not correspond to the formula for $T_G(\delta)$
 derived in the same paper.
\end{thebibliography}
\end{document}